\documentclass{jpp}

\usepackage{amssymb,amsmath,framed}
\usepackage{graphicx}
\usepackage{epstopdf, epsfig}
\usepackage[unicode=true,pdfusetitle,bookmarks=true,bookmarksnumbered=false,bookmarksopen=false,breaklinks=false,pdfborder={0 0 0},backref=false,colorlinks=true]{hyperref}
\hypersetup{citecolor=blue,filecolor=blue,linkcolor=blue,urlcolor=blue}
\usepackage[dvipsnames]{xcolor}

\makeatletter
\@ifpackageloaded{stix} 
{
}{                      
	\usepackage{mathrsfs}
  \usepackage{newtxtext,newtxmath} 	

  \DeclareFontFamily{OML}{txmi1}{}
  \DeclareFontShape{OML}{txmi1}{m}{it}{<->txmi1}{}
  \DeclareSymbolFont{myletter}{OML}{txmi1}{m}{it}
  \DeclareMathSymbol{v}{\mathalpha}{myletter}{`v}

}
\ifx\bm\undefined
  \newcommand{\bm}[1]{{\boldsymbol {\mathrm #1}}} 
\else
  \renewcommand{\bm}[1]{{\boldsymbol {\mathrm #1}}} 
\fi
\makeatother

\newcommand{\f}[2]{\frac{#1}{#2}}
\newcommand{\mr}[1]{\mathrm{#1}}
\newcommand{\lf}{\left}
\newcommand{\ri}{\right}

\newcommand{\dd}[2]{\frac{\rmd#1}{\rmd#2}}
\newcommand{\pp}[2]{\frac{\p #1}{\p #2}}

\newcommand{\Div}{\nabla\cdot}
\newcommand{\Curl}{\nabla\times}
\newcommand{\nbl}{\nabla}

\newcommand{\+}{{\perp}}
\newcommand{\unit}[1]{\hat{\bm{#1}}}

\makeatletter
\newcommand{\vast}{\bBigg@{4}}
\newcommand{\Vast}{\bBigg@{5}}

\makeatother

\newcommand{\calB}{\mathcal{B}}

\newcommand{\calO}{\mathcal{O}}

\newcommand{\calU}{\mathcal{U}}

\DeclareMathAlphabet{\mathpzc}{OT1}{pzc}{m}{it}

\newcommand{\rmd}{\mathrm{d}}
\newcommand{\rme}{\mathrm{e}}

\newcommand{\rmi}{\mathrm{i}}

\newcommand{\rmA}{\mathrm{A}}

\newcommand{\rmH}{\mathrm{H}}

\newcommand{\rmS}{\mathrm{S}}

\shorttitle{Relativistic Hall Reduced MHD}
\shortauthor{Y. Kawazura}

\title{Hall magnetohydrodynamics in a relativistically strong mean magnetic field}

\author{
  Y. Kawazura\aff{1,2}\corresp{\email{kawazura@tohoku.ac.jp}}
}

\affiliation{
  \aff{1}Frontier Research Institute for Interdisciplinary Sciences, Tohoku University, 6-3 Aoba, Aramaki, Sendai 980-8578, Japan
  \aff{2}Department of Geophysics, Graduate School of Science, Tohoku University, 6-3 Aoba, Aramaki, Aoba-ku, Sendai 980-8578 Japan
}

\date{\today}

\begin{document}

\maketitle

\begin{abstract}
  This Letter presents a magnetohydrodynamic model that describes the small-amplitude fluctuations with wavelengths comparable to ion inertial length in the presence of a relativistically strong mean magnetic field.
  The set of derived equations is virtually identical to the non-relativistic Hall reduced magnetohydrodynamics~\citep{Schekochihin2019}, differing only by a few constants that take into account the relativistic corrections.
  This means that all the properties of kinetic Alfv\'en turbulence and ion cyclotron turbulence inherent in the non-relativistic Hall regime persist unchanged even in a magnetically dominated regime.
\end{abstract}

\section{Introduction}
Turbulence of relativistically magnetized plasmas (here defined as the magnetic field energy exceeding the rest mass energy of particles) can be found in a number of astrophysical systems, e.g., pulsar and black hole magnetospheres, coronae of accretion disks, and jets from active galactic nuclei.
The turbulent fluctuations of magnetic field in these systems can be dissipated and converted into the thermal and nonthermal energy of particles~\citep{Zhdankin2017a,Comisso2018b,Zhdankin2019,Nattila2022}, which are potential sources of bright electromagnetic radiation we observe on the earth.
As the reservoir of magnetic energy is huge, even small-amplitude fluctuations give rise to significant heating and acceleration.
Thus, understanding the properties of turbulent fluctuations in relativistically magnetized plasmas (also known as magnetically dominated plasmas) is one of the most important themes in modern high-energy astrophysics.

In the vast majority of studies of relativistic turbulence, either ideal magnetohydrodynamics (MHD) or fully kinetic Vlasov-Maxwell equations are used (with a few exceptions that used the resistive MHD to describe proper reconnection~\citep[e.g.,][]{Ripperda2020,Ripperda2022} and that used relativistic Braginskii equations to include weakly collisional effects~\citep[e.g.,][]{Foucart2016,Foucart2017}).
Ideal MHD is, by definition, only able to describe the large-scale dynamics, and thus, it is not suitable for studying the dissipation of fluctuations, which usually occurs on scales smaller than the ion inertial length or the ion Larmor radius.
Vlasov-Maxwell equations, on the other hand, can properly describe the dissipation processes, but it is a rather too complex model.  
In fact, when Vlasov-Maxwell equations are used for relativistic turbulence, they are solved only by means of \textit{ab initio} particle-in-cell simulations, while the analytical study of small-scale physics of relativistic plasmas is underdeveloped. 
The aim of this study is to formulate a comprehensive and useful relativistic magnetohydrodynamic model that is valid even at small scales.

The extension of relativistic ideal MHD to incorporate the small-scale effects was first proposed by \citet{Koide2009} (which is then rederived using a variational principle~\citep{Kawazura2017a}). 
The model is often called the relativistic extended MHD (XMHD) which includes the Hall effect, the rest mass inertia of electrons, and the thermal inertia of electrons.
When the inertia effects of electrons are neglected, this set of equations is referred to as a relativistic Hall MHD (HMHD).
To date, relativistic XMHD and relativistic HMHD have been widely used, e.g., for studying magnetic reconnection~\citep{Comisso2014,Comisso2018a,Asenjo2019,Yang2019a,Yang2019b}, magnetofluid topological connection~\citep{Asenjo2015a, Asenjo2015b,Comisso2020}, and linear wave propagation~\citep{Kawazura2017b,Kawazura2022c}.
However, these models have not been used for turbulence\footnote{Note that the non-relativistic version of XMHD~\citep{Kimura2014,Abdelhamid2015} was used for turbulence of the solar wind~\citep{Abdelhamid2016b}.}.
Since the set of relativistic XMHD equations is much more complicated than that of non-relativistic XMHD or relativistic ideal MHD, it may be too difficult to solve relativistic XMHD as it is, even using direct numerical simulations. 
Alternatively, in this work, we reduce relativistic HMHD to make it more tractable by assuming the presence of a mean magnetic field --- a technique commonly used for non-relativistic models.

When the spatial scale of the turbulent fluctuations is much smaller than the scale of energy injection (which is macroscopic in many astrophysical systems), the large-scale magnetic field effectively behaves like a mean field for the fluctuations~\citep{Kraichnan1965,Howes2008c}.
Therefore, as the turbulent cascade progresses, the fluctuations become smaller amplitude and more elongated along the mean field.
Consequently, the ideal MHD asymptotically becomes reduced MHD (RMHD)~\citep{Kadomtsev1974,Strauss1976}.
While non-relativistic RMHD has been widely used in studies of magnetically confined fusion, solar wind~\citep[e.g.,][]{Chen2011}, planetary magnetospheres~\citep[e.g.,][]{Watanabe2010}, and accretion flows~\citep{Kawazura2022a}, relativistic RMHD was formulated only recently~\citep{Chandran2018,TenBarge2021}.
Remarkably, relativistic RMHD and non-relativistic RMHD are formally identical except for the definition of Alfv\'en speed which is modified such that it never exceeds the speed of light.
This means that all the properties of turbulence described by non-relativistic RMHD are true even in the relativistic regime (for example, the Alfv\'en and slow waves are energetically decoupled~\citep{Schekochihin2009} while the fast waves are entirely ordered out; see \citet{TenBarge2021} for a detailed discussion on relativistic RMHD).

In the non-relativistic regime, the same reduction procedure can be adopted for Hall MHD, and the resulting model is called Hall reduced MHD (HRMHD)~\citep{Gomez2008}, which is valid at the ion inertial length.
HRMHD can also be derived by gyrokinetics when ions are cold, and the electron beta is order unity~\defcitealias{Schekochihin2019}{S19}\citep[][hereafter S19]{Schekochihin2019}.
The Alfv\'en waves and slow waves (which are decoupled in RMHD) are reorganized into the kinetic Alfv\'en waves (KAW) and oblique ion cyclotron waves (ICW) in HRMHD.

Now, it is quite natural to ask whether the relativistic effects alter the properties of KAW and ICW turbulence in the HRMHD limit.
Here, we formulate the relativistic version of HRMHD (i.e., the relativistic extension of \citetalias{Schekochihin2019}, or equivalently, the inclusion of Hall effect in \citet{Chandran2018,TenBarge2021}).
This is a simple and comprehensive model that is valid at the microscopic scales when the background magnetic field is relativistically strong.  
We find that the relativistic HRMHD is almost identical to the non-relativistic HRMHD~\citepalias{Schekochihin2019}, and thus the properties of KAW and ICW in a non-relativistic regime are also true in a magnetically dominated regime.

\section{Derivation of relativistic HRMHD}
Consider quasi-neutral relativistic ion and electron fluids with infinitely small electron-to-ion mass ratio.
To describe the time evolution of such plasmas, we use relativistic HMHD~\citep{Kawazura2017b}, which consists of the mass conservation law
\begin{subequations}
\begin{equation}
  \pp{}{t}\lf( n\gamma \ri) + \Div\lf( n\gamma\bm{u} \ri) = 0,
	\label{e:HMHD continuity}
\end{equation}
the momentum equation
\begin{equation}
  \pp{}{t}\lf( nh\gamma^2\bm{u} \ri) + \Div\lf( nh\gamma^2\bm{u}\bm{u} \ri) = -c^2\nbl p + c^2\rho_q\bm{E} + c\bm{J}\times\bm{B},
	\label{e:HMHD e.o.m.} 
\end{equation}
the generalized Ohm's law
\begin{equation}
  \bm{E} + \f{\bm{u}}{c}\times\bm{B} = \f{1}{\gamma e n}\lf( \rho_q \bm{E} + \f{\bm{J}}{c}\times\bm{B} - \nbl p_\rme \ri),
	\label{e:HMHD Ohm's law}
\end{equation}
and Maxwell's equations
\begin{align}
  \Div\bm{E} = 4\pi\rho_q,
	\label{e:Maxwell div E} \\
  -\pp{\bm{E}}{t} + c\Curl\bm{B} = 4\pi\bm{J},
	\label{e:Maxwell curl B} \\
	\Div\bm{B} = 0,
	\label{e:Maxwell div B} \\
  \pp{\bm{B}}{t} + c\Curl\bm{E} = 0,
	\label{e:Maxwell curl E} 
\end{align}
\end{subequations}
where $e$ is the elementary charge, $c$ is the speed of light, $n$ is the rest frame number density, $h$ is the total thermal enthalpy, $p$ is the total thermal pressure, $p_\rme$ is the thermal pressure of electrons, $\bm{u}$ is the flow velocity, $\gamma = 1/\sqrt{1 - |\bm{u}|^2/c^2}$ is the Lorentz factor, $\bm{J}$ is the electric current, $\rho_q$ is the charge density, $\bm{E}$ is the electric field, and $\bm{B}$ is the magnetic field.
The relativistic ideal MHD is recovered when the right-hand side of \eqref{e:HMHD Ohm's law} is neglected.

In what follows, we assume that all fields are separable into spatio-temporally constant background (symbols with a subscript 0) and fluctuations (symbols with $\delta$ in front), viz. $n = n_0 + \delta n, \; \bm{B} = (B_0 + \delta B_\|)\unit{z} + \delta\bm{B}_\+$, and so on. 
Electrons are assumed to be isothermal, i.e., $\delta p_\rme = T_{\rme0}\delta n$, where $T_{\rme0}$ is the background electron temperature.
Here, $\unit{z} = \bm{B}_0/|\bm{B}_0|$, and $\| (\+)$ denotes the parallel (perpendicular) component to $\bm{B}_0$.
We also assume that the mean flow is absent, i.e. $\bm{u}_0 = 0$.
Plugging the constant background fields into \eqref{e:HMHD Ohm's law}, \eqref{e:Maxwell div E}, and \eqref{e:Maxwell curl B}, one finds $\bm{E}_0 = \bm{J}_0 = \rho_{q0} = 0$.
Then, we impose the reduced MHD ordering 
\begin{equation}
  \f{\delta n}{n_0} \sim \f{\delta \bm{B}}{B_0} \sim \f{\bm{u}}{v_\rmA} \sim \f{\delta p_\rme}{p_{\rme0}} \sim \f{k_\|}{k_\+} \sim \epsilon \ll 1, \quad \pp{}{t} \sim \omega \sim k_\| v_\rmA,
\end{equation}
where $\omega$ and $\bm{k}$ are the frequency and wavenumber of the fluctuations, respectively.
Here, we have defined the relativistic Alfv\'en speed.
\begin{equation}
  v_\rmA = \f{cB_0}{\sqrt{4\pi n_0 h_0 + B_0^2}} = \sqrt{\f{\sigma}{1 + \sigma}}\,c,
  \label{e:def v_A}
\end{equation}
where $\sigma = B_0^2/4\pi n_0 h_0$ is the magnetization parameter.
When $\sigma \gtrsim 1$ (equivalently $v_\rmA \approx c$), the plasma is relativistically magnetized.
Since $\bm{u}$ is small, the relativistic effect of bulk flow is absent, i.e., $\gamma \approx 1$, but this is acceptable because we are interested in the microscopic scales where bulk flow is generally small while the thermal energy and/or magnetic energy can be relativistic.
We also assume that the ions are cold while the electron can be relativistically hot,
\begin{equation}
  T_{\rmi0} \ll T_{\rme0} \ll \sqrt{\f{m_\rmi}{m_\rme}}\,m_\rme c^2 \approx 20\,\mr{MeV},
   \label{e:HMHD ordering -1-}
\end{equation}
where $T_{s0}$ and $m_s$ are temperature and mass of the species $s$.
The upper bound for the electron temperature is required so that the relativistic thermal inertia of electrons is negligible at the ion inertial scale.
Since the ions are cold, the thermal inertia of the ions is also negligible, i.e. $h_0 \approx m_\rmi c^2$. 
These conditions enforces the restriction,
\begin{equation}
  \beta_\rme = \lf(\f{2T_{\rme0}}{m_\rmi c^2}\ri)\f{1}{\sigma} \ll \sqrt{\f{m_\rme}{m_\rmi}},
\end{equation}
where $\beta_\rme = 8\pi p_{\rme0}/B_0^2$ is the electron beta. 
This is different from the assumptions $T_{\rmi0} \ll T_{\rme0}$ and $\beta_\rme \sim 1$ that are used in the non-relativistic HRMHD~\citepalias{Schekochihin2019} because they cannot be satisfied when the magnetization is relativistic, i.e., $\sigma \gtrsim 1$.  
In this work, we consider the electron beta to be lower than that of non-relativistic HMHD, which is formally allowed as long as $\epsilon \ll \beta_\rme$.
In other words, we treat $\beta_\rme$ as order unity, although it is much smaller than $\sqrt{m_\rme/m_\rmi}$, because $\epsilon$ is assumed to be even smaller than $\beta_\rme$.
However, $\beta_\rme$ may not be too small because when $\beta_\rme \sim m_\rme/m_\rmi$, electron rest mass inertia becomes non-negligible, and thus the HMHD approximation breaks down~\citep{Zocco2011}.
To summarize, we assume the range of $\beta_\rme$ and $\epsilon$ as
\begin{equation}
 \epsilon \sim  \f{m_\rme}{m_\rmi} \ll \beta_\rme \ll \sqrt{\f{m_\rme}{m_\rmi}}.
 \label{e:HMHD ordering -2-}
\end{equation}

We, then, follow the derivation of non-relativistic RMHD by \citet{Schekochihin2009}.
We explicitly keep $v_\rmA/c$ and $k_\+ d_\rmi$ in the ordering so that one can take the non-relativistic and/or long-wavelength limit simply by neglecting the corresponding terms.
First, we adopt the expansion $\bm{u} = \bm{u}^{(1)} + \bm{u}^{(2)} + \calO(\epsilon^3 v_\rmA)$ and substitute it into~\eqref{e:HMHD continuity}.
The $\calO(\epsilon^0 n_0 \omega)$ terms yield
\begin{equation}
  \nbl_\+\cdot\bm{u}_\+^{(1)} = 0.
\end{equation}
This allows us use a stream function leading to $\bm{u}_\+^{(1)} = \unit{z}\times\nbl_\+\Phi$, where $\Phi = (c/B_0)\phi$, and $\phi$ is the electrostatic potential.
From the $\calO(\epsilon^1 n_0 \omega)$ terms, one obtains
\begin{equation}
  \Bigg( \pp{}{t} + \bm{u}_\+^{(1)}\cdot\nbl_\+ \Bigg)\f{\delta n}{n_0} = -\Bigg( \nbl_\+\cdot\bm{u}_\+^{(2)} + \pp{u_\|^{(1)}}{z} \Bigg).
	\label{e:HMHD continuity 1st order}
\end{equation}
From the lowest order terms in \eqref{e:Maxwell div B}, one can use a magnetic flux function leading to $v_\rmA(\delta \bm{B}_\+^{(1)}/B_0) = \unit{z}\times\nbl_\+\Psi$, where $\Psi = -(v_\rmA/B_0)A_\|$, and $A_\|$ is the parallel component of the vector potential.
The $\calO(\epsilon^0\omega)$ terms in \eqref{e:HMHD e.o.m.} give the pressure balance
\begin{align}
  \f{B_0}{4\pi}\delta B_\| = -\delta p_\rme = -T_{\rme0}\delta n.
  \label{e:pressure balance}
\end{align}
Up to this point, the derivation is the same as the non-relativistic RMHD.

Next, we expand $\bm{E} = \bm{E}^{(1)} + \bm{E}^{(2)} + \calO(\epsilon^3 v_\rmA B_0/c)$. 
Using electromagnetic potentials, the first and second order terms become
\begin{equation}
  \bm{E}^{(1)} = -\nbl_\+\phi, \quad \bm{E}^{(2)} = -\f{1}{c}\pp{\bm{A}}{t} - \unit{z}\pp{\phi}{z}.
  \label{e:E expanded}
\end{equation}
We also expand $\bm{J} = \bm{J}^{(1)} + \bm{J}^{(2)} + \calO(\epsilon^3 ck_\+B_0)$, plug it into \eqref{e:HMHD Ohm's law}, and remove $\rho_q$ using \eqref{e:Maxwell div E} to yield
\begin{multline}
  \f{c}{v_\rmA B_0}\Big( \underbrace{\bm{E}^{(1)}}_{\sim\epsilon} + \underbrace{\bm{E}^{(2)}}_{\sim\epsilon^2} \Big) + \f{c}{v_\rmA}\bigg( \underbrace{\f{\delta n}{n_0}}_{\sim\epsilon^2} - \underbrace{\f{\nbl_\+\cdot\bm{E}_\+^{(1)}}{4\pi e n_0}}_{\sim\epsilon^2(k_\+d_\rmi)(v_\rmA/c)^2} \bigg)\f{\bm{E}^{(1)}}{B_0} + \f{1}{v_\rmA}\bigg( \underbrace{\bm{u}^{(1)}}_{\sim\epsilon} - \underbrace{\f{\bm{J}^{(1)}}{en_0}}_{\sim\epsilon (k_\+ d_\rmi)} \bigg)\times\unit{z} \\
  + \f{1}{v_\rmA}\bigg( \underbrace{\bm{u}^{(1)}}_{\sim\epsilon^2} - \underbrace{\f{\bm{J}^{(1)}}{en_0}}_{\sim\epsilon^2 (k_\+ d_\rmi)} \bigg)\times\f{\delta \bm{B}}{B_0} + \f{1}{v_\rmA}\bigg[ \underbrace{\lf( \f{\delta n}{n_0} \ri)\bm{u}^{(1)} + \bm{u}^{(2)}}_{\sim\epsilon^2} + \underbrace{\lf( \f{\delta n}{n_0} \ri)\f{\bm{J}^{(1)}}{en_0} - \f{\bm{J}^{(2)}}{en_0}}_{\sim\epsilon^2 (k_\+ d_\rmi)} \bigg]\times\unit{z} \\
  + \f{c}{v_\rmA en_0 B_0}\bigg[\underbrace{\nbl_\+\delta p_\rme}_{\sim\epsilon\beta_\rme (k_\+d_\rmi)} + \underbrace{\lf( \f{\delta n}{n_0} \ri)\nbl_\+\delta p_\rme + \unit{z}\pp{}{z}\delta p_\rme}_{\sim\epsilon^2\beta_\rme (k_\+d_\rmi)}\bigg] = 0,
  \label{e: HMHD Ohm's law -1-}
\end{multline}
where $d_\rmi = \sqrt{m_\rmi c^2/4\pi n_0 e^2}$ is the ion inertial length.
As we mentioned above, $\beta_\rme$ is assumed to be $\calO(\epsilon^0)$ because $\epsilon \ll \beta_\rme$.
We then manipulate \eqref{e:Maxwell curl B} to get
\begin{multline}
  \f{4\pi\bm{J}}{cB_0} = -\f{1}{c^2}\unit{z}\times\pp{}{t}\bigg( \underbrace{\bm{u}_\+^{(1)}}_{\sim \epsilon^2 k_\+(v_\rmA/c)^2} - \underbrace{\f{\bm{J}_\+^{(1)}}{en_0}}_{\sim \epsilon^2 k_\+(k_\+ d_\rmi)(v_\rmA/c)^2} \bigg) + \underbrace{\f{1}{ce n_0 B_0}\pp{}{t}\nbl_\+\delta p_\rme}_{\sim \epsilon^2 \beta_\rme k_\+(k_\+ d_\rmi)(v_\rmA/c)^2} \\
  + \underbrace{\unit{z}\times\pp{}{z}\f{\delta \bm{B}_\+}{B_0}}_{\sim \epsilon^2 k_\+} + \underbrace{\nbl_\+\bigg( \f{\delta B_\|}{B_0} \bigg)\times\unit{z}}_{\sim \epsilon k_\+} + \underbrace{\nbl_\+ \times\f{\delta\bm{B}_\+}{B_0}}_{\sim \epsilon k_\+} + \calO(\epsilon^3 k_\+).
\end{multline}
Collecting the terms order-by-order, one obtains
\begin{equation}
\begin{aligned}
  &\bm{J}_\+^{(1)} = \f{c}{4\pi}\nbl_\+ \delta B_\| \times \unit{z},\quad J_\|^{(1)} = \f{c}{4\pi}\unit{z}\cdot(\nbl_\+\times\delta\bm{B}_\+),\\ 
  &\bm{J}_\+^{(2)} = \f{c}{4\pi}\lf[ -\f{B_0}{c^2}\unit{z}\times\pp{}{t}\lf( \bm{u}_\+^{(1)} - \f{\bm{J}_\+^{(1)}}{en_0} \ri) + \f{1}{cen_0}\pp{}{t}\nbl_\+\delta p_\rme + \unit{z}\times\pp{}{z}\delta\bm{B}_\+ \ri], \quad J_\|^{(2)} = 0.
  \label{e:J 1st and 2nd order}
\end{aligned}
\end{equation}
The terms including $\p/\p t$ in $\bm{J}_\+^{(2)}$ are originated from the displacement current, which disappears in the non-relativistic limit, and all the other terms are the same as the non-relativistic case.
Plugging $\bm{J}_\+^{(1)}$ into the $\calO(\epsilon)$ terms in \eqref{e: HMHD Ohm's law -1-}, one obtains the pressure balance \eqref{e:pressure balance} again, where the isothermal electrons are assumed. 
Then, we can further manipulate $\bm{J}_\+^{(2)}$ as
\begin{equation}
  \bm{J}_\+^{(2)} = \f{c}{4\pi}\lf( \f{B_0}{c^2}\pp{}{t}\nbl_\+\Phi + \unit{z}\times\pp{}{z}\delta\bm{B}_\+ \ri).
  \label{e:J 2nd order}
\end{equation}
This is notable because the Hall term and the electron pressure gradient term, both of which becomes finite at the $d_\rmi$ scale, are canceled, and therefore, the electric current (up to the second order) is identical to that of relativistic RMHD.
The reason for this cancellation is rather straightforward. 
We evaluate the electric field $\bm{E}$ in the displacement current by using Ohm's law, which is the momentum equation of electrons. 
Therefore, the pressure balance cancels some of the terms of the electric current exactly in the same way as the first order of the momentum equation in the relativistic RMHD (which is identical to \eqref{e:RHRMHD Phi -1-} shown below).

Next, we substitute \eqref{e:E expanded} and \eqref{e:J 1st and 2nd order} into \eqref{e:HMHD e.o.m.} to get the equations for $\Phi$ and $u_\|$.
As we found in the previous paragraph, both $\bm{E}$ and $\bm{J}$ do not contain corrections due to the Hall effect, meaning that \eqref{e:HMHD e.o.m.} ends up with the same equations as the relativistic RMHD.
The $z$ component of $\calO(\epsilon^1\omega)$ terms give
\begin{equation}
  \pp{u_\|}{t} + \lf\{ \Phi, u_\| \ri\} = (1 + \sigma)v_\rmA^2\lf[ \pp{}{z}\lf( \f{\delta B_\|}{B_0} \ri) + \f{1}{v_\rmA}\lf\{ \Psi, \f{\delta B_\|}{B_0}\ri\} \ri],
  \label{e:RHRMHD U -1-}
\end{equation}
while multiplying $\unit{z}\cdot\nbl_\+\times$ to $\calO(\epsilon^1\omega)$ terms gives
\begin{equation}
  \pp{}{t}\nbl_\+^2\Phi + \lf\{ \Phi, \nbl_\+^2\Phi \ri\} = v_\rmA\lf( \pp{}{z}\nbl_\+^2\Psi + \f{1}{v_\rmA}\lf\{ \Psi, \nbl_\+^2\Psi \ri\} \ri), 
  \label{e:RHRMHD Phi -1-}
\end{equation}
where $\lf\{ f, g \ri\} = (\p_x f)(\p_y g) - (\p_x g)(\p_y f)$.

Next, we derive the equations for $\Psi$ and $\delta B_\|$.
This time, unlike the momentum equation, there will be the corrections due to the Hall effect and electron pressure gradient.
The $\calO(\epsilon^2)$ terms in \eqref{e: HMHD Ohm's law -1-} are
\begin{multline}
  \f{\bm{E}^{(2)}}{B_0} + \bigg(\f{\delta n}{n_0} - \f{\nbl_\+\cdot\bm{E}_\+^{(1)}}{4\pi e n_0} \bigg)\f{\bm{E}^{(1)}}{B_0} + \f{1}{c}\bigg( \bm{u}^{(1)} - \f{\bm{J}^{(1)}}{en_0} \bigg)\times\f{\delta \bm{B}}{B_0} \\ + \f{1}{c}\bigg[ \lf( \f{\delta n}{n_0} \ri)\bm{u}^{(1)} + \bm{u}^{(2)} - \f{\bm{J}^{(2)}}{en_0} \bigg]\times\unit{z} + \f{1}{en_0 B_0}\bigg(\unit{z}\pp{}{z}\delta p_\rme\bigg) = 0.
  \label{e: HMHD Ohm's law -2-}
\end{multline}
Note that a few terms have been canceled out using the pressure balance \eqref{e:pressure balance}.
The $z$-component of \eqref{e: HMHD Ohm's law -2-} yields
\begin{equation}
  \pp{\Psi}{t} = v_\rmA\lf[ \pp{}{z}\lf( \Phi + \sqrt{(1 + \sigma)}\, d_\rmi v_\rmA\f{\delta B_\|}{B_0} \ri) + \f{1}{v_\rmA}\lf\{ \Psi,\, \Phi + \sqrt{(1 + \sigma)}\, d_\rmi v_\rmA\f{\delta B_\|}{B_0} \ri\} \ri].
  \label{e:RHRMHD Psi -1-}
\end{equation}
Meanwhile, \eqref{e:HMHD continuity 1st order} and \eqref{e:pressure balance} are rearranged as
\begin{align}
  \nbl_\+\cdot\bm{u}_\+^{(2)} + \pp{u_\|^{(1)}}{z} = \f{\sigma c^2}{c_\rmS^2}\dd{}{t}\lf( \f{\delta B_\|}{B_0} \ri),
  \label{e:div u}
\end{align}
where $c_\rmS = \sqrt{(n_{\rmi0}T_{\rme0})/(n_{\rme0}m_i)} \ll (m_\rme/m_\rmi)^{1/4}c$ is the sound speed,
and \eqref{e:J 1st and 2nd order} and \eqref{e:J 2nd order} are rearranged as
\begin{align}
  \f{1}{en_0}\lf( \nbl_\+\cdot\bm{J}_\+^{(2)} + \pp{J_\|^{(1)}}{z} \ri) = \f{\sqrt{(1 + \sigma)}\, d_\rmi v_\rmA}{c^2}\pp{}{t}\nbl_\+^2\Phi.
  \label{e:div J}
\end{align}
Note that the right hand side of \eqref{e:div J} equals to zero in the non-relativistic limit (i.e., $\Div\bm{J} = 0$).
Combining \eqref{e:div u} and \eqref{e:div J} with $\unit{z}\cdot\nbl_\+\times$ to \eqref{e: HMHD Ohm's law -2-} yields
\begin{multline}
  -\lf( 1 + \f{\sigma c^2}{c_\rmS^2} \ri)\lf[ \pp{}{t}\lf( \f{\delta B_\|}{B_0} \ri) + \lf\{ \Phi,\, \f{\delta B_\|}{B_0} \ri\} \ri] + \f{\sqrt{(1 + \sigma)}\, d_\rmi v_\rmA}{c^2}\lf( \pp{}{t}\nbl_\+^2\Phi + \lf\{ \Phi,\, \nbl_\+^2\Phi \ri\} \ri) \\
  + \pp{u_\|}{z} + \f{1}{v_\rmA}\lf\{ \Psi,\, u_\| \ri\} - \sqrt{(1 + \sigma)}\, d_\rmi \lf( \pp{}{z}\nbl_\+^2\Psi + \f{1}{v_\rmA}\lf\{ \Psi,\, \nbl_\+^2\Psi \ri\} \ri) = 0.
  \label{e:RHRMHD B -1-}
\end{multline}
The time derivative of $\nbl_\+^2\Phi$ can be eliminated by substituting \eqref{e:RHRMHD Phi -1-}, which results in
\begin{multline}
  \lf( 1 + \f{\sigma c^2}{c_\rmS^2} \ri)\lf[ \pp{}{t}\lf( \f{\delta B_\|}{B_0} \ri) + \lf\{ \Phi,\, \f{\delta B_\|}{B_0} \ri\} \ri] \\
  = \pp{u_\|}{z} + \f{1}{v_\rmA}\lf\{ \Psi,\, u_\| \ri\} - \sqrt{(1 + \sigma)}\, d_\rmi\lf( 1 - \f{v_\rmA^2}{c^2} \ri)\lf( \pp{}{z} + \f{1}{v_\rmA^2}\lf\{ \Psi,\, \nbl_\+^2\Psi \ri\} \ri).
  \label{e:RHRMHD B -2-}
\end{multline}

Finally, \eqref{e:RHRMHD U -1-}, \eqref{e:RHRMHD Phi -1-}, \eqref{e:RHRMHD Psi -1-}, and \eqref{e:RHRMHD B -2-} are reorganized into
\begin{subequations}
\begin{align}
  &\dd{}{t}\nbl_\+^2\Phi = v_\rmA\nbl_\|\nbl_\+^2\Psi,
  \label{e:RHRMHD Phi -*-} \\
  &\pp{\Psi}{t} = v_\rmA\nbl_\|\lf( \Phi + v_\rmA\rho_\rmH\calB \ri),
  \label{e:RHRMHD Psi -*-} \\
  &\dd{\calU}{t} = v_\rmS\nbl_\|\calB,
  \label{e:RHRMHD U -*-} \\
  &\dd{\calB}{t} = \nbl_\|\lf( v_\rmS\calU - \rho_\rmH \nbl_\+^2\Psi \ri),
  \label{e:RHRMHD B -*-} 
\end{align}
\end{subequations}
where
\begin{subequations}
\begin{align}
  &\dd{}{t} = \pp{}{t} + \bm{u}_\+\cdot\nbl_\+ = \pp{}{t} + \lf\{ \Phi, \cdots \ri\},\quad \nbl_\| = \pp{}{z} + \f{\delta\bm{B}_\+}{B_0}\cdot\nbl_\+ = \pp{}{z} + \f{1}{v_\rmA} \lf\{ \Psi, \cdots \ri\}, \\
  &\calU = \f{u_\|}{v_\rmA}, \quad \calB = \sqrt{(1 + \sigma)\lf[ 1 + (1 + \sigma)v_\rmA^2/c_\rmS^2 \ri]}\,\f{\delta B_\|}{B_0},\\
  &v_\rmA = \sqrt{\f{\sigma}{1 + \sigma}}\,c, \quad v_\rmS = \sqrt{\f{\sigma}{1 + (1 + \sigma)v_\rmA^2/c_\rmS^2}}\,c, \quad \rho_\rmH = \sqrt{\f{1}{1 + (1 + \sigma)v_\rmA^2/c_\rmS^2}}\,d_\rmi.
  \label{e:RHRMHD consts}
\end{align}
\end{subequations}
Note that $v_\rmA^2/c_\rmS^2 \approx 2/\beta_\rme$ when $T_{\rmi0}/T_{\rme0} \ll 1$.
We find that \eqref{e:RHRMHD Phi -*-}-\eqref{e:RHRMHD B -*-} are identical to non-relativistic HRMHD~\citepalias[(5.14)-(5.17) of ][]{Schekochihin2019}), except for the definition of constants~\eqref{e:RHRMHD consts} which become those of \citetalias{Schekochihin2019} in the limit of $\sigma \to 0$.
Thereby, all the physical properties of the non-relativistic HRMHD (e.g., the conservation of energy and helicity~\citepalias[(5.24) and (5.68) of ][]{Schekochihin2019} and the linear dispersion relation~\citepalias[(5.26) of ][]{Schekochihin2019}) are valid even in a magnetically dominated regime $\sigma \gtrsim 1$.
Meanwhile, one retrieves the relativistic RMHD by taking the limit of $\rho_\rmH \to 0$.
Note that RMHD of \citet{Chandran2018,TenBarge2021} are written in the Els\"asser variables while \eqref{e:RHRMHD Phi -*-}-\eqref{e:RHRMHD B -*-} are not, as is shown by \citet{Galtier2006} that introducing the Els\"asser for HMHD is complicated.

There are three important facts that make relativistic HRMHD and non-relativistic HRMHD formally identical.
The first is that the electric current of relativistic HRMHD is identical to that of relativistic RMHD due to the electron pressure balance (see \eqref{e:J 1st and 2nd order}).
The second is that the displacement current only appears in the second order of the electric current.
The third is that the Lorentz force due to the displacement current (i.e., $\bm{J}^{(2)}\times\bm{B}_0$) is the time derivative of $\bm{E}^{(1)}\times \bm{B}_0$ drift, which happens to be the left-hand side of the momentum equation \eqref{e:HMHD e.o.m.} and balances the non-relativistic Lorentz force $\bm{J}^{(1)}\times\delta \bm{B}$ in the right-hand side (given the pressure balance). 
Thus, the relativistic Lorentz force ends up with merely proportional to the non-relativistic Lorentz force, and its prefactor does not include $d_\rmi$.

Lastly, one finds that the Hall transition scale $\rho_\rmH$ becomes smaller and vanishes eventually as the magnetization $\sigma$ increases, meaning that the fluctuations behaves like those of RMHD at $d_\rmi$ scale when the magnetization is strong.
This behavior was first discovered by \citet{Kawazura2017b}, and the reason for it is that the Hall transition happens at the scale of $\sim v_\rmA/\Omega_\rmi$ (which is equal to $d_\rmi$ in the non-relativistic limit) where $\Omega_\rmi$ is ion cyclotron frequency, and this scale becomes smaller as the magnetization becomes larger because $v_\rmA \to c$ and $\Omega_\rmi \to \infty$ as $\sigma \to \infty$.

\section{Discussion}
In this Letter, we have shown that the properties of fluctuations in the sub-Hall transition scale ($k_\+ \rho_\rmH \gtrsim 1$) found by non-relativistic HRMHD remain the same even when the mean magnetic field is relativistically strong and/or the electron temperature is relativistically hot (but much less than $(m_\rmi/m_\rme)^{1/2}m_\rme c^2$).
More specifically, for example, \citetalias{Schekochihin2019} theoretically showed that the Alfv\'enic and compressive cascades in the RMHD range ($k_\+ \rho_\rmH \ll 1$) are rearranged into KAW and ICW cascades below the Hall transition scale($k_\+ \rho_\rmH \gg 1$), and the scalings of KAW cascade are $k_\+^{-7/3}$ for the magnetic energy and $k_\+^{-13/3}$ for the kinetic energy while those of ICW cascade are $k_\+^{-5/3}$ for the magnetic energy and $k_\+^{-11/3}$ for the kinetic energy.
These scalings are consistent with the numerical simulation of non-relativistic HMHD~\citep{Meyrand2012}.
We find that the same scalings apply to the relativistic regime, although there have been no simulation of turbulence in relativistic ion and electron plasmas that elucidated the scalings at the transition scale.
Note that the relativistic PIC simulation of ion and electron plasma turbulence found the presence of spectral break at the ion Larmor scale~\citep{Zhdankin2019}.
Furthermore, \citetalias{Schekochihin2019} also showed that the KAW and ICW cascades eventually turn into electron and ion heating respectively, and we have shown that this scenario of energy partition between ions and electrons is also true in the relativistic regime. 

In closing, we admit that the ordering assumptions we have made in this study, namely \eqref{e:HMHD ordering -1-} and \eqref{e:HMHD ordering -2-}, are rather restrictive and may not be directly relevant to realistic astrophysical objects. 
However, even though HMHD is consistent with kinetic theory only for the cold ion limit~\citep{Ito2004,Howes2009}, the simulations of turbulence at the transition scale via HMHD and through the hybrid particle-in-cell with $T_\rmi = T_\rme$ demonstrated nearly the same results~\citep{Papini2019}, suggesting that the predictions made by HMHD are practically useful beyond its theoretical limitations.
We think the same holds true for our relativistic HRMHD.
%
%
 

\section*{Acknowledgements}
The author thanks Shigeo Kimura for fruitful conversations.
This work was supported by JSPS KAKENHI grant JP20K14509. 

\section*{Declaration of interests}
The author reports no conflict of interest.


\bibliographystyle{jpp}
\bibliography{../../references}

\end{document}